\documentclass[onecolumn,showpacs,preprintnumbers,amsmath,amssymb]{revtex4}

\usepackage{graphicx}
\usepackage{dcolumn}
\usepackage{bm}

\newcommand{\kB}{k_\mathrm{B}}
\newcommand{\muB}{\mu_\mathrm{B}}
\DeclareMathOperator{\erf}{erf}
\DeclareMathOperator{\erfc}{erfc}
\DeclareMathOperator{\PG}{P}
\providecommand{\abs}[1]{\lvert#1\rvert}

\begin{document}

\preprint{}

\title{Theory of Evaporative Cooling with Energy-Dependent Elastic
Scattering Cross Section and Application to Metastable Helium}

\author{Paul J. J. Tol, Wim Hogervorst and Wim Vassen}

\affiliation{%
Laser Centre Vrije Universiteit, De Boelelaan 1081, 1081 HV
Amsterdam, The Netherlands}

\date{\today}

\begin{abstract}
The kinetic theory of evaporative cooling developed by Luiten {\em et
al.} [Phys.\ Rev.\ A $\bm{53}$, 381 (1996)] is extended to include the
dependence of the elastic scattering cross section on collision
energy. We introduce a simple approximation by which the transition
range between the low-temperature limit and the unitarity limit is
described as well. Applying the modified theory to our measurements on
evaporative cooling of metastable helium we find a scattering length
$\abs{a}=10(5)$~nm.
\end{abstract}


\maketitle


\section{Introduction}

Evaporative cooling is at present the most powerful method to increase
the phase-space density of a dilute trapped gas in order to reach
Bose-Einstein condensation (BEC) or Fermi degeneracy
\cite{ketterle1996}. The atoms are first laser-cooled to temperatures
of 1 mK or lower and trapped either magnetically in a Ioffe quadrupole
(IQ) trap, or optically in a dipole trap. Here the energy of the atoms
is redistributed by elastic collisions. A few atoms acquiring energies
larger than the trap depth $\varepsilon_\text{t}$ (with a maximum
energy of $2 \varepsilon_\text{t}$) are expelled from the trap. This
reduces significantly the mean energy of the trapped ensemble while
only a relatively small number of atoms is lost. In forced evaporative
cooling the trap depth is gradually lowered, during which cooling
increases the density and the elastic collision rate, accelerating
evaporation.

In the Walraven group kinetic theory has been developed to describe
this evaporative cooling process for cold atomic gases \cite{luiten,
pinkse1998}. The theory assumes collisions occurring in the s-wave
limit, an energy-independent elastic scattering cross section $\sigma$
and sufficient ergodicity in the trap. A truncated Boltzmann
distribution is introduced and shown to be consistent with the
classical Boltzmann equation applied to a trap of finite depth. Then
equations of statistical mechanics are obtained describing the
kinetics, particle loss and energy loss. Closed expressions are
derived that are directly applicable to an IQ trap. However, the
assumption of an energy-independent cross section $\sigma$ is not
justified when evaporation starts in the transition range between the
low-temperature limit and the unitarity limit. This is the case, for
instance, in metastable helium, where evaporative cooling typically
starts at $\sim$1~mK \cite{robert,pereira2001,herschbach2003}.  In
order to understand evaporative cooling under these conditions as
well, this paper extends the theory to incorporate an energy-dependent
$\sigma$. The theory is subsequently applied to interpret our
evaporative cooling experiments with metastable helium, allowing the
s-wave scattering length $a$ to be extracted. As BEC has been realized
by two groups in France \cite{robert, pereira2001}, $a$ is known to be
large and positive. Published experimental values are $a=+20(10)$~nm
\cite{robert} and $a=+16(8)$~nm \cite{pereira2001}, both obtained from
Bose-condensed clouds. Recent experiments in Paris \cite{leduc} with
clouds just above the BEC transition suggest that $a$ is smaller than
previously published values. Theoretical values determined from
precisely calculated molecular potentials are +8.3~nm \cite{starck}
and +12.2~nm \cite{dickinson, gadea}.

This paper is organized as follows. In Section \ref{secTheory} the
theory of evaporative cooling based on the papers of Luiten {\em et
al.}~\cite{luiten} and Pinkse {\em et al.}~\cite{pinkse1998} is
summarized and extended. First, Section \ref{secDepth} gives the
density and energy of an atomic cloud in a magnetic trap of finite
depth and Section \ref{secEvap} provides the equations necessary to
calculate all relevant loss rates during the cooling process, assuming
$\sigma$ is energy-independent. Then, in Section \ref{secSigma}, the
atom and energy loss rates due to evaporation are recalculated, now
with an energy-dependent cross section $\sigma$. Additionally, a simple
approximate expression is presented for an effective cross section
that can be implemented directly in the equations for the evaporation
loss rates given before in Section \ref{secEvap}. The theory is used
in Section \ref{secSweep} to interpret evaporative cooling experiments
in metastable helium performed in Amsterdam and a value for the
scattering length is deduced. Section \ref{secSummary} summarizes and
compares the extracted scattering length with results obtained in the
other experiments and with theoretical values from molecular potential
calculations.

\section{Theory}
\label{secTheory}

\subsection{Density in magnetic traps with finite depth}
\label{secDepth}

The trapping potential due to a magnetic field $B(\bm{r})$ with
(local) minimum ${B_0}$ is given by $U(\bm{r})=m_J g_J \muB
[B(\bm{r})-{B_0}]$, where $g_J$ is the Land\'e $g$-factor and $m_J$
the magnetic quantum number of the state of the trapped atoms. We use
metastable helium in the 2~$^3$S$_1$, $m_J=1$ state with $g_J \approx
2$. The trapping potential of an IQ trap can be approximated by
\begin{equation}
U(x,y,z)=\sqrt{\alpha^2(x^2+y^2)+(U_0+\beta z^2)^2}-U_0\,,
\label{leafgravpot}
\end{equation}
with minimum potential energy
\begin{equation}
{U_0}=m_J g_J \muB {B_0} \,,
\label{leafgravpotzero}
\end{equation}
where the effect of gravity is neglected.

According to Luiten {\em et al.}\ \cite{luiten} the atoms in a
magnetic trap with a finite depth are described well by a Boltzmann
energy distribution that is truncated at an energy equal to the trap
depth. The phase-space distribution is assumed to be only a function
of the energy of the atoms. At low temperatures this is the case if
there are many elastic collisions between the atoms. At higher
temperatures in the IQ trap, the atoms occupy a part of space large
enough that higher-order terms of the potential break the axial
symmetry of Eq.~(\ref{leafgravpot}) and the motion of the atoms
becomes ergodic even without collisions. Below a summary is given
of the equations needed to calculate the density distribution in a
magnetic trap \cite{luiten}.

The (thermal) density distribution in an infinitely deep trap is
\begin{equation}
{n_{\infty }}(\bm{r})={n_0}\exp\biggl[-\frac{U(\bm{r})}{\kB T}\biggr]\,.
\end{equation}
When only atoms with an energy smaller than $\varepsilon_\text{t}$ are
trapped, the density distribution becomes
\begin{equation}
n(\bm{r})=\PG\bigl[\tfrac{3}{2},\kappa (\bm{r})\bigr]{n_{\infty }}(\bm{r})\,,
\end{equation}
with incomplete gamma function $\PG$ \cite{gammafunction} and
\begin{equation}
\kappa (\bm{r})=
\left\{\begin{aligned}
&\frac{\varepsilon_\text{t}-U(\bm{r})}{\kB T}\,, &U(\bm{r}) \le
\varepsilon_\text{t}\,, \\
&0\,, &U(\bm{r}) > \varepsilon_\text{t}\,.
\end{aligned}\right.
\end{equation}
In this truncated Boltzmann distribution $T$ is still called the
temperature, although strictly speaking a thermodynamic temperature is
not defined for a nonequilibrium distribution. The central density is
given by $n(0) = \PG\left[\frac{3}{2},\kappa (\bm{r}) \right]n_0$;
parameter ${n_0}$ is equal to the central density only in the limit of
infinite trap depth. It is defined by
\begin{equation}
{n_0}=N/{V_\text{e}}\,,
\label{NVe}
\end{equation}
with number of trapped atoms $N$ and reference volume $V_\text{e}$,
which is equal to the effective volume $N/n(0)$ in the limit of a deep
trap. In general the reference volume is given by
\begin{equation}
{V_\text{e}}={{\varLambda }^3}\zeta \,,
\label{Ve}
\end{equation}
with thermal de Broglie wavelength
\begin{equation}
\varLambda =\sqrt{ \frac{2 \pi \hbar^2}{m \kB T} }
\end{equation}
and trap-dependent single-atom partition function $\zeta$.  For an IQ
trap
\begin{equation}
\begin{aligned}
&\zeta =\zeta _{\infty }^{0}\bigl[\PG(4,\eta
)+\tfrac{2}{3}\tfrac{{U_0}}{\kB T}\PG(3,\eta )\bigr] \,, \\
&\zeta _{\infty }^{0}=6{{{A_{\text{IQ}}}(\kB  T)}^4}, \\
&{A_{\text{IQ}}}={m^{3/2}}\big/\bigl(4{\textstyle\sqrt{2\beta}}\,{{\alpha
}^2}{{\hbar }^3}\bigr) \,,
\end{aligned} 
\label{zetaleaf}
\end{equation}
with truncation parameter 
\begin{equation}
\eta =\frac{\varepsilon_\text{t}}{\kB T}\,.
\end{equation}
The total internal energy of the trapped atoms is
\begin{equation}
E=\frac{12\PG(5,\eta)+6\frac{U_0}{\kB T}\PG(4,\eta)}{3\PG(4,\eta) +
2\frac{U_0}{\kB T}\PG(3,\eta)}N \kB T\,.
\label{totalenergy}
\end{equation}
Although ${n_0}$ is not the central density and a thermodynamic
temperature cannot be given for this nonequilibrium distribution, the
(central) phase-space density (or degeneracy parameter) is still
${n_0}{{\varLambda }^3}$.

\subsection{Evaporative cooling with constant cross section}
\label{secEvap}

The behavior of a cloud of atoms during an rf sweep is simulated with
the model of Luiten {\em et al.}\ \cite{luiten} and Pinkse {\em et
al.}\ \cite{pinkse1998}. The version of the model described here
assumes that every atom with an energy greater than the trap depth
$\varepsilon_\text{t}$ leaves the trap (three-dimensional
evaporation). Their trajectories should bring these atoms sufficiently
fast to the exit area of the trap (one of the saddle points or, when
an rf field is applied, positions where the rf field is resonant with
the magnetic field), so they are removed without first colliding with
another atom. Another assumption is that collisions occur in the
s-wave regime.

At the start, $N$ atoms at temperature $T$ are contained in a trap
with depth $\varepsilon_\text{t}$ determined by the magnetic field
configuration.  After a time step of negligible size the rf power is
turned on: the rf frequency $\omega_\text{rf}$, the time dependence of
which is known beforehand, determines the truncation energy
$\varepsilon_\text{t}= \hbar {{\omega }_{\text{rf}}}-U_0$. Then in
small steps (for our situation 10~ms) the loss of atoms and energy due
to inelastic collisions, trap changes (spilling) and evaporation is
determined, as discussed below. The temperature is found by solving
numerically the equation for the total energy of the trapped atoms,
Eq.~(\ref{totalenergy}).  If experimentally an extra temperature
increase is found, for instance due to instability of the power
supplies, this can be added separately. The calculation is stopped
when either no atoms are left or when BEC is reached. In the last case
the phase-space density is ${n_0}{{\varLambda }^3}\geq
{g_{3/2}}(1)\approx 2.6$, with polylogarithm function
${g_n}(z)=\sum_{k=1}^{\infty}{z^k}/{k^n}$.

References \cite{luiten,pinkse1998} do not give all loss rates in a
suitable form, so first some intermediate equations are given and the
energy density is derived. Trapped atoms have energy
\begin{equation}
\varepsilon (\bm{r},\bm{p})=U(\bm{r})+{p^2}/2m\,.
\label{atomenergy}
\end{equation}
The number of atoms with energy between $\varepsilon$ and $\varepsilon
+\,\text{d} \varepsilon$ is $\rho
(\varepsilon)f(\varepsilon)\,\text{d} \varepsilon$, with phase-space
distribution
\begin{equation}
f(\varepsilon)={n_0}{{\varLambda }^3}\exp\Bigl(-\frac{\varepsilon}{\kB
T}\Bigr)
\end{equation}
and (in an IQ trap) an energy density of states
\begin{equation}
\rho (\varepsilon) = {A_{\text{IQ}}}\bigl({{\varepsilon}^3} +
2{U_0}{{\varepsilon}^2}\bigr)\,.
\end{equation}
The phase-space distribution can be given as a function of $\bm{r}$
and $\bm{p}$ via Eq.~(\ref{atomenergy}) and is normalized so that the
total number of trapped atoms is
\begin{equation}
N=\frac{1}{{{(2\pi \hbar )}^3}}\iint
f(\bm{r},\bm{p})\,\text{d}^3r\,\text{d}^3p\,,
\end{equation}
where the integration is done over the volume in phase space where
$\varepsilon\leq \varepsilon_\text{t}$. The density is given by
\begin{equation}
\begin{split}
n(\bm{r})&=\frac{1}{{{(2\pi  \hbar )}^3}}\int f(\bm{r},\bm{p})\,\text{d}^3p \\
&=\frac{{n_0}{{\varLambda }^3}}{{{(2\pi \hbar )}^3}}\int
_{0}^{{\sqrt{2m[\varepsilon_\text{t}-U(\bm{r})]}}}
\exp\biggl(-\frac{U(\bm{r})+{p^2}/2m}{\kB T}\biggr)4\pi
{p^2}\,\text{d} p \\
&=\PG\bigl[\tfrac{3}{2},\kappa (\bm{r})
\bigr]{n_0}\exp\biggl(-\frac{U(\bm{r})}{\kB T}\biggr)\,.
\end{split}
\end{equation}
Similarly, the energy density is
\begin{equation}
\begin{split}
e(\bm{r})&=\frac{1}{{{(2\pi \hbar )}^3}}\int \varepsilon
(\bm{r},\bm{p})f(\bm{r},\bm{p})\,\text{d}^3p \\
&=\bigl\{\tfrac{3}{2}\kB T \PG\bigl[\tfrac{5}{2},\kappa
(\bm{r})\bigr]+U(\bm{r}) \PG\bigl[\tfrac{3}{2},\kappa
(\bm{r})\bigr]\bigr\}{n_0}\exp\biggl(-\frac{U(\bm{r})}{\kB
T}\biggr)\,.
\end{split}
\end{equation}

\subsubsection{Inelastic collisions}

Inelastic collisions occur with background gas atoms, between pairs of
trapped atoms and between three trapped atoms. Corresponding loss
rates are
\begin{align}
{\dot{N}_{\text{bgr}}}&=-\frac{1}{\tau }\int
n(\bm{r})\,\text{d}^3r=-N/\tau\,, \\
{\dot{N}_{\text{2b}}}&=-G\int
{n^2}(\bm{r})\,\text{d}^3r\,, \\
{\dot{N}_{\text{3b}}}&=-L\int
{n^3}(\bm{r})\,\text{d}^3r\,,
\end{align}
respectively; corresponding energy loss rates are
\begin{align}
{\dot{E}_{\text{bgr}}}&=-\frac{1}{\tau }\int
e(\bm{r})\,\text{d}^3r=-E/\tau\,, \\
{\dot{E}_{\text{2b}}}&=-G\int
e(\bm{r})n(\bm{r})\,\text{d}^3r\,, \\
{\dot{E}_{\text{3b}}}&=-L\int
e(\bm{r}){n^2}(\bm{r})\,\text{d}^3r\,,
\end{align}
respectively. A dot denotes a derivative with respect to time. The
integrals have to be calculated numerically for every time step. The
constant $\tau $ is the lifetime of the trap, determined
experimentally. Fedichev {\em et al.}\
\cite{fedichev1996a,fedichev1996b} have calculated the other two
constants for spin-polarized metastable helium.  For $T<0.1$~mK, $G$
is only dependent on magnetic field. The maximum value is $6 \times
10^{-13}$~cm$^3$/s at 750~G, but as the contribution of two-body
collisions is more significant towards the end of the sweep when the
cloud is concentrated at the center, the value for $B\leq10$~G, which
is $G=3 \times 10^{-14}$~cm$^3$/s, can be used
\cite{fedichev1996a}. The three-body loss-rate constant is given by
$L=11.6\hbar {a^4}/m$, with scattering length $a$
\cite{fedichev1996b}.

\subsubsection{Spilling}

When the trap shape or depth is changed, the eigenstates of the
trapping potential with highest energies can become unbound. Atoms in
these states are spilled from the trap. This process does not depend
on collisions; when the potential is changed only by lowering the
truncation energy $\varepsilon_\text{t}$, spilling does not alter
parameters $T$ and ${n_0}$. After instantaneous lowering of the trap
depth from $\varepsilon_\text{t}$ to $\varepsilon_\text{t}'$, the
change in the number of atoms due to spilling is $\Delta {N_{\theta
}}=-\int_{\varepsilon_\text{t}'}^{\varepsilon_\text{t}}\rho
(\varepsilon) f(\varepsilon) \,\text{d} \varepsilon$ and the
corresponding change in energy is $\Delta {E_{\theta }}=-\int
_{\varepsilon_\text{t}'}^{{\varepsilon}_\text{t}}\varepsilon\rho
(\varepsilon) f(\varepsilon) \,\text{d} \varepsilon$. With
${n_0}{{\varLambda }^3}=N/\zeta $ and Eq.~(\ref{zetaleaf}),
integration yields
\begin{align}
\Delta {N_{\theta }}&=-N \left[1-\frac{3\PG(4,{{\eta
}'})+2\frac{{U_0}}{\kB T}\PG(3,{{\eta }'})}{3\PG(4,\eta
)+2\frac{{U_0}}{\kB T}\PG(3,\eta )}\right]=-{N}\biggl[1-\frac{\zeta
({{\eta }'})}{\zeta (\eta )}\biggr]\,,
\label{Nspill} \\[1ex]
\Delta {E_{\theta }}&=-N \kB T \frac{12\PG(5,\eta )+6\frac{{U_0}}{\kB 
T}\PG(4,\eta )-12\PG(5,{{\eta }'})-6\frac{{U_0}}{\kB 
T}\PG(4,{{\eta }'})}{3\PG(4,\eta )+2\frac{{U_0}}{\kB T}\PG(3,\eta
)}\,,
\end{align}
where ${{\eta }'}=\varepsilon_\text{t}'/\kB T\leq \eta $; the values
of $N$ and $T$ are those before the step takes place.

\subsubsection{Evaporation}

After an elastic collision of two trapped atoms with energy
$\varepsilon <\varepsilon_\text{t}$, one atom may have an energy
$\varepsilon>\varepsilon_\text{t}$ and leave the trap. After this
thermal escape, or evaporation, the average energy per atom has become
smaller. The number and energy loss rates due to evaporation are
\begin{align}
{\dot{N}_{\text{ev}}}&=-{\sqrt{\frac{8\kB T}{\pi
m}}}\,n_{0}^{2}\sigma {{\text{e} }^{-\eta }}{V_{\text{ev}}}\,,
\label{Nev} \\
{\dot{E}_{\text{ev}}}&={\dot{N}_{\text{ev}}}\left(\eta
+\frac{{W_{\text{ev}}}}{{V_{\text{ev}}}}\right)\kB T\,,
\label{Eev}
\end{align}
with effective volumes for evaporation
\begin{align}
{V_{\text{ev}}}&={{\varLambda }^3}\zeta _{\infty
}^{0}\left\{\Bigl(1+\tfrac{2}{3}\tfrac{{U_0}}{\kB T}\Bigr)\left[\eta
-\textstyle\sum _{i=1}^{4}\PG(i,\eta )\right]-\PG(5,\eta )\right\}\,, \\
{W_{\text{ev}}}&={{\varLambda }^3}\zeta _{\infty
}^{0}\left\{\Bigl(1+\tfrac{2}{3}\tfrac{{U_0}}{\kB T}\Bigr)\left[\eta
-\textstyle\sum _{i=1}^{5}\PG(i,\eta )\right]-\PG(6,\eta )\right\}\,,
\end{align}
and elastic scattering cross section $\sigma$. The model has been
developed for an energy-independent cross section $\sigma =8\pi
{a^2}$.

\subsection{Incorporation of an energy-dependent cross section}
\label{secSigma}

The model as described in the previous section is restricted to the
low-temperature limit, where $\sigma$ is energy-independent. In this
section the theory will be adapted to include energy-dependence,
although remaining in the s-wave regime.  In our experiment the
temperature at the start of the rf sweep is about 1~mK, where the
dependence of $\sigma$ on the relative velocity of the colliding atoms
cannot be neglected. For the derivation of atom loss rate
${\dot{N}_{\text{ev}}}$ and energy loss rate ${\dot{E}_{\text{ev}}}$
with an energy-dependent cross section $\sigma$, Luiten's derivation
\cite{jomthesis} is adapted. The elastic collision event rate
${{\varGamma }_\text{\!c}}$, which is the number of collisions per
second occurring in the whole cloud, is also derived. The average
collision rate per atom is $2{{\varGamma }_\text{\!c}}/N$. This rate
is not needed in the sweep simulation, but it is an interesting
quantity, related to the atom loss rate ${\dot{N}_{\text{ev}}}$, which
is the collision event rate under the condition that afterwards one of
the atoms has enough energy to leave the trap.

\subsubsection{Collision event rate}

\begin{figure}
\includegraphics{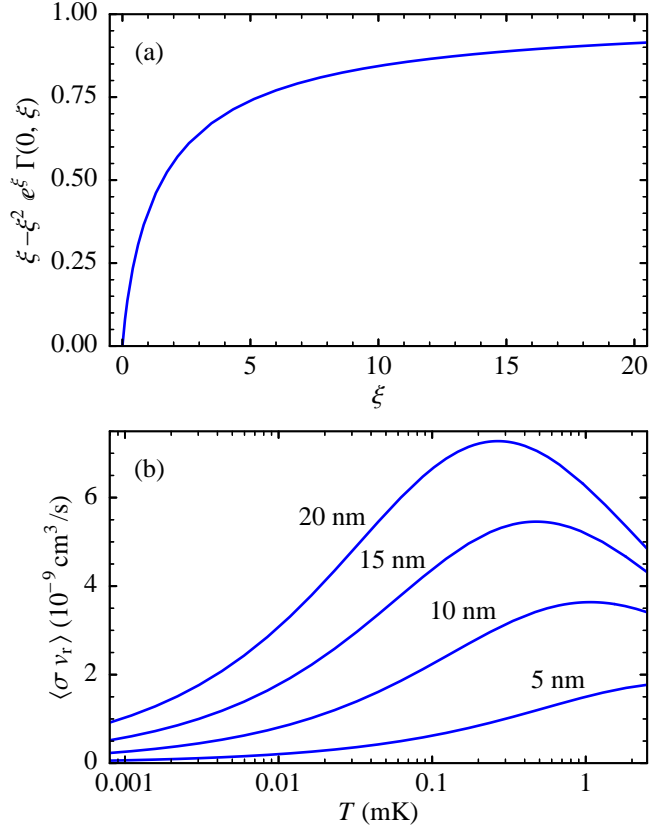}
\caption{(a) Factor $\langle \sigma v_\text{r} \rangle / (8 \pi a^2
\langle v_\text{r} \rangle)$ as a function of $\xi$, and (b) the rate
coefficient as a function of temperature for metastable helium assuming
four different values for the scattering length $a$.}
\label{scatrate}
\end{figure}

First the collision rate in the case of an (untruncated) Boltzmann
velocity distribution is examined. The two-body elastic collision
event rate is
\begin{equation}
{{\varGamma }_\text{\!c}}=\int \tfrac{1}{2}{{n(\bm{r})}^2}\langle \sigma
{v_\text{r}}\rangle
\,\text{d}^3r\,,
\end{equation}
where rate coefficient $\langle \sigma {v_\text{r}}\rangle $ is the
product of cross section $\sigma $ and relative velocity
${v_\text{r}}$ of the colliding pair of atoms, averaged over the
thermal velocity distribution
\begin{displaymath}
\frac{1}{2{\sqrt{\pi }}}{{\biggl(\frac{m}{\kB T}\biggr)}^{3/2}}
v_\text{r}^{2}\exp\biggl(-\frac{m v_\text{r}^{2}}{4 \kB T}\biggr)\,\text{d}
{v_\text{r}}\,.
\end{displaymath}
Assuming the rate coefficient is not dependent on position due to the
magnetic field, the collision event rate can be written as
\begin{equation}
{{\varGamma
}_\text{\!c}}=\frac{N}{2}\frac{{V_{\text{2e}}}}{{V_\text{e}}}{n_0}\langle
\sigma {v_\text{r}}\rangle
\label{gaussgamma}
\end{equation}
and the average collision rate per atom as
$({V_{\text{2e}}}/{V_\text{e}}){n_0}\langle \sigma {v_\text{r}}\rangle
$, with
\begin{align}
{V_\text{e}}&=\int \frac{n(\bm{r})}{n_0}\,\text{d}^3r=N/{n_0}\,, \\
{V_{\text{2e}}}&=\int {{\left(\frac{n(\bm{r})}{n_0}\right)}^2}\,\text{d}^3r\,.
\end{align}
For a Gaussian shaped density distribution
${V_{\text{2e}}}/{V_\text{e}}=1/{\sqrt{8}}$, for the density
distribution in an IQ trap with $U_0=0$ (and infinite $\eta$)
${V_{\text{2e}}}/{V_\text{e}}=1/{\sqrt{32}}$. The collision rate at
position $\bm{r}$ is $n(\bm{r})\langle \sigma {v_\text{r}}\rangle
$. Using the approximation $\sigma = 8 \pi a^2 / (k^2 a^2 + 1)$ with
thermal wave vector $k=m {v_\text{r}}/(2\hbar )$, we get
\begin{equation}
\langle \sigma {v_\text{r}}\rangle =8\pi {a^{2 }}\langle
{v_\text{r}}\rangle \bigl[\xi_\text{c}-{\xi_{\text{c}}^2}
{{\text{e} }^{{{\xi }_\text{c}}}}\Gamma (0,{{\xi }_\text{c}})\bigr]\,,
\label{colrate}
\end{equation}
with shorthand notation ${{\xi }_\text{c}}={{\hbar }^2}/({a^2}m \kB
T)$ and average relative velocity $\langle {v_\text{r}}\rangle
=4{\sqrt{\kB T/(\pi m)}}$. Figure~\ref{scatrate} shows the quantity in
square brackets and the temperature dependence of the rate coefficient
in the case of metastable helium. In the low-temperature limit
${{\text{lim}}_{\xi \rightarrow \infty }} \bigl[\xi-\xi^2 {{\text{e}
}^{\xi }}\Gamma (0,\xi )\bigr]=1$.

For a truncated Boltzmann distribution the collision event rate can be
written as
\begin{equation}
{{\varGamma }_\text{\!c}}=\frac{1}{2{{m(2\pi \hbar )}^6}}\iiint \sigma
 \bigl(\abs{{\bm{p}_2} - {\bm{p}_1}}\bigr)
 \,\abs{{\bm{p}_2}-{\bm{p}_1}}
 \,f(\bm{r},{\bm{p}_1})f(\bm{r},{\bm{p}_2})\,\text{d}^3{p_1}
 \,\text{d}^3{p_2}\,\text{d}^3r\,.
\end{equation}
The integration is over all $(\bm{r},{\bm{p}_i})$ with
$U(\bm{r})+p_i^{2}/2m\leq \varepsilon_\text{t}$ for both $i=1$ and
$i=2$. A more convenient coordinate system uses average momentum
$\bm{P}=({\bm{p}_1}+{\bm{p}_2})/2$ and relative momentum
$\bm{q}={\bm{p}_2}-{\bm{p}_1}$, with angle $\theta$ between $\bm{P}$
and $\bm{q}$ and azimuthal angle $\varphi$. With Jacobian
$\,\text{d}^3{p_1} \,\text{d}^3{p_2}=\,\text{d}^3P\,\text{d}^3q$ we
get
\begin{equation}
{{\varGamma }_\text{\!c}}=\frac{n_{0}^{2}{{\varLambda }^6}}{2{{m(2\pi \hbar
)}^6}}\int {{\text{e} }^{-2U/\kB T}}\int \sigma (q)\, q
{{\text{e} }^{-{q^2}/4m \kB T}}\int {{\text{e}
}^{-{P^2}/m \kB T}}\,\text{d}^3P\,\text{d}^3q\,\text{d}^3r\,,
\label{generalgamma}
\end{equation}
where the integration range is now given by
\begin{equation}
{P^2}+\frac{{q^2}}{4}+P q \abs{\cos \theta}\leq
2m\bigl[\varepsilon_\text{t}-U(\bm{r})\bigr]\,.
\label{gammacond}
\end{equation}
To eliminate the angle dependence, the integrand is multiplied by the
fraction $F$ of colliding pairs at position $\bm{r}$ for fixed $P$ and
$q$ that are part of the truncated Boltzmann distribution:
\begin{equation}
F=\frac{1}{4\pi }\int _{0}^{2\pi }\int \sin \theta
\,\text{d} \theta \,\text{d} \varphi\,,
\end{equation}
where the integration over $\theta$ is restricted by condition
(\ref{gammacond}). The result is
\begin{equation}
F=\frac{{Q^2}-{P^2}-{q^2}/4}{P q}\,,
\end{equation}
with $Q={\sqrt{2m[\varepsilon_\text{t}-U(\bm{r})]}}$ the maximum
momentum an atom can have at position $\bm{r}$ without leaving the
trap.  This fraction has to be restricted to the physical range
between 0 and 1. For $P>({Q^2}-{q^2}/4)^{1/2}$ at least one of the
colliding atoms would have more momentum than is possible at position
$\bm{r}$ and the fraction is zero. For $0<P<Q-q/2$ all angles $\theta
$ are possible and the fraction is one. The integration over $\bm{P}$
is therefore divided into two parts:
\begin{multline}
{{\varGamma }_\text{\!c}}=\frac{n_{0}^{2}{{\varLambda }^6}}{2{{m(2\pi \hbar
)}^6}}\int {{\text{e} }^{-2U/\kB T}}\int _{0}^{2Q}\sigma(q)\, q
{{\text{e} }^{-{q^2}/4m \kB T}} \\
\Biggl(\int _{0}^{Q-q/2}
{{\text{e} }^{-{P^2}/m \kB T}}\,\text{d}^3P+\int
_{Q-q/2}^{{\sqrt{Q^2-q^2/4}}}F {{\text{e} }^{-{P^2}/m
\kB T}}\,\text{d}^3P\Biggr)\,\text{d}^3q\,\text{d}^3r\,.
\label{eventratetwoparts}
\end{multline}
For an IQ trap [Eq.~(\ref{leafgravpot})], the integration over
position can be made one-dimensional \cite{luiten} using
\begin{equation}
\int \mathcal{F} [U(\bm{r})]\,\text{d}^3r=\frac{4\pi }{{{\alpha
}^2}{\sqrt{\beta }}}\int
_{0}^{{\varepsilon}_\text{t}}{\sqrt{U}}(U+{U_0})\mathcal{F}
(U)\,\text{d} U \,.
\end{equation}
Introducing scaled variables $y=q/{\sqrt{m \kB T}}$ and $\kappa
=(\varepsilon_\text{t}-U)/\kB T$, and scaled constant $\eta
=\varepsilon_\text{t}/\kB T$ the collision event rate after
integration over $\bm{P}$ becomes
\begin{multline}
\varGamma_\text{\!c}=\frac{{{n_0^2}{\text{e} }^{-2\eta
}}}{{{\alpha }^2} {\sqrt{\beta m}}}{{(\kB T)}^3}\int
_{0}^{\eta }\int _{0}^{{\sqrt{8\kappa }}} \sigma(y) {\sqrt{\eta
-\kappa }}\, \Bigl(\eta -\kappa +\tfrac{U_0}{\kB
T}\Bigr)\, {y^2} \\
\biggl[2-2\exp {\Bigl({\sqrt{2\kappa }}
\,y-\tfrac{1}{2}y^2\Bigr)}+ {\sqrt{\pi }}\, y \exp {\Bigl(2 \kappa
-\tfrac{1}{4}y^2\Bigr)} \erf{\Bigl({\sqrt{2\kappa
}}-\tfrac{1}{2}y\Bigr)}\biggr]\,\text{d} y \,\text{d} \kappa \,.
\label{biggamma}
\end{multline}

For an energy-dependent cross section $\sigma$ with $\hbar k=q/2$ this
equation has to be integrated numerically. For a constant cross
section, after integration over $y$, the expression for
$\varGamma_\text{\!c}$ contains terms
\begin{equation}
\int _{0}^{\eta }{{(\eta -\kappa )}^x}{{\text{e} }^{l \kappa
}}\erfc{\sqrt{l \kappa }}\,\text{d} \kappa =\Gamma (x+1){{\text{e}
}^{l \eta }} \bigl[\PG(x+1,l \eta )-\PG\bigl(x+\tfrac{3}{2},l \eta
\bigr)\bigr]\big/{l^{x+1}}
\end{equation}
and
\begin{multline}
\int _{0}^{\eta }{{(\eta -\kappa )}^x}\kappa\, {{\text{e} }^{\kappa
}}\erfc{\sqrt{\kappa }}\,\text{d} \kappa =\Gamma (x+1){{\text{e}
}^{\eta }}\bigl\{(\eta +1)\bigl[\PG(x+2,\eta
)-\PG\bigl(x+\tfrac{7}{2},\eta \bigr)\bigr] \\
-\tfrac{3}{2} \PG\bigl(x+\tfrac{5}{2},\eta \bigr)-(x+2)
 \PG(x+3,\eta)+\bigl(x+\tfrac{7}{2}\bigr) \PG\bigl(x+\tfrac{9}{2},\eta
 \bigr)\bigr\} \,,
\end{multline}
with Euler gamma function $\Gamma(z)=\Gamma(z,0)$ and complementary
error function $\erfc(z)=1-\erf(z)$, which has the property
\cite{abramowitz}
\begin{equation}
{{\text{e} }^{\kappa
}}\erfc{\sqrt{\kappa }}=\frac{{\sqrt{\kappa }}}{\pi }\int
_{0}^{\infty }\frac{{{\text{e} }^{-t}}}{{\sqrt{t}}\,(t+\kappa
)}\,\text{d} t \,.
\end{equation}
The collision event rate for constant $\sigma$ becomes
\begin{multline}
\varGamma_\text{\!c}=
\tfrac{1}{2\sqrt{2}}
n_0^2 \sigma \sqrt{\tfrac{\kB T}{\pi m}} \varLambda^3 \zeta_\infty
\Bigl\{ 1+\tfrac{4}{3}\tfrac{U_0}{\kB T}
-8 \text{e}^{-\eta} \Bigl( \eta-1+\tfrac{2}{3}\tfrac{U_0}{\kB T}\eta \Bigr) \\
+ \tfrac{1}{3} \text{e}^{-2\eta}
\Bigl[ \eta^4 + 4\eta^3 + 6\eta^2 - 6\eta - 27 +
\tfrac{4}{3}\tfrac{U_0}{\kB T}
\bigl( 2\eta^3 + 6\eta^2 + 6\eta - 3 \bigr) \Bigr] \Bigr\} \,.
\end{multline}

When $\sigma$ is energy-dependent, $\varGamma_\text{\!c}$ needs to be
calculated numerically. However, in the limit of large $\eta$ the
velocity distribution becomes thermal and Eq.~(\ref{biggamma}) is
expected to produce the same result as Eq.~(\ref{gaussgamma}). For
common values of $\eta$ and a cross section given by $\sigma = 8 \pi
a^2 / (k^2 a^2 + 1)$, Eqs.~(\ref{gaussgamma}) and (\ref{colrate}) can
be used as an approximation: comparing exact numerical calculations of
${{\varGamma }_\text{\!c}}\big/\bigl(\frac{N}{2}
\frac{{V_{\text{2e}}}} {{V_\text{e}}} {n_0}\bigr)$ with the rate
coefficient as given by Eq.~(\ref{colrate}), at $\eta =10$ deviations
are smaller than 0.1\% and at $\eta =5$ smaller than 6\% (for
$m<100$~u, $\abs{a}<100$~nm, $U_0/\muB<100$~G, $T<5$~mK).

\subsubsection{Evaporation rates}

The collision event rate $\varGamma_\text{\!c}$ is given by
Eq.~(\ref{generalgamma}), with constraints to the integration range
incorporated by a factor $F$ in the integrand. The atom loss rate
${\dot{N}_{\text{ev}}}$ is given by the same equation, but with an
extra factor for the fraction of collisions for which afterwards one
of the atoms has enough energy to leave the trap. Since $P$ and $q$
are the same before and after collisions and all scattering angles
${{\theta }'}$ have equal probability in the s-wave regime, this
factor is $(1-F)$. In the case of the collision event rate, the
integration over $P$ had to be divided into two parts [see
Eq.~(\ref{eventratetwoparts})]. Here only one part is left, as the
integrand including $F(1-F)$ with $F=1$ is zero. The evaporation rate
becomes
\begin{multline}
\dot{N}_{\text{ev}}=
-\frac{4 \pi n_0^2 \varLambda^6}{\alpha^2 \sqrt{\beta}\, 2m(2 \pi \hbar)^6}
\int_{0}^{\varepsilon_\text{t}}
{\sqrt{U}}(U+{U_0}){{\text{e} }^{-2U/\kB T}}
\int _{0}^{2Q} \sigma(q)\, 4\pi {q^3} {\text{e}}^{-{q^2}/4m \kB T} \\
\int_{Q-q/2}^{\sqrt{Q^2-q^2/4}}
F(1-F)\,4\pi{P^2}{\text{e}}^{-{P^2}/m \kB T}\,\text{d}
P\,\text{d} q\,\text{d} U\,,
\label{generalNev}
\end{multline}
and after integration over $P$:
\begin{multline}
\dot{N}_{\text{ev}}= -\frac{n_0^2
{\text{e}}^{-2\eta}}{8\alpha^2\sqrt{\beta m}} (\kB T)^3 \int_0^\eta
\int_0^{\sqrt{8\kappa}}
\sigma(y)\,\sqrt{\eta-\kappa}\,\Bigl(\eta-\kappa+\tfrac{U_0}{\kB
T}\Bigr)\,y \\
\biggl\{
-2\bigl[ y(2+y^2-8\kappa) + 2\sqrt{2\kappa}\,(6+y^2-8\kappa) \bigr]
\exp{\bigl( \sqrt{2\kappa}\,y-\tfrac{1}{2}y^2 \bigr)} \\
+2 \bigl[8y + \sqrt{8\kappa-y^2}\,(6+y^2-8\kappa) \bigr]
-\sqrt{\pi}\,\bigl[12+4y^2+y^4-16(2+y^2)\kappa+64\kappa^2\bigr] \\
\exp{\bigl( 2\kappa-\tfrac{1}{4}y^2 \bigr)}
\Bigl[\erf{\Bigl( \sqrt{2\kappa-\tfrac{1}{4}y^2} \,\Bigr)}
- \erf{\Bigl( \sqrt{2\kappa}-\tfrac{1}{2}y \Bigr)} \Bigr]
\biggr\} \,\text{d}y \,\text{d}\kappa \,.
\label{bigNev}
\end{multline}

For the energy loss rate ${\dot{E}_{\text{ev}}}$ there is an extra
factor in the integrand for the energy of the lost atom,
\begin{equation}
{{\varepsilon}_4}=U+\frac{1}{2m}\biggl({P^2}+\frac{{q^2}}{4}+P q
\abs{\cos {{\theta }'}}\biggr)\,.
\end{equation}
This means ${\dot{E}_{\text{ev}}}$ is given by Eq.~(\ref{generalNev})
with the factor $(1-F)$ replaced by
\begin{equation}
G=\frac{1}{4\pi }\int _{0}^{2\pi }\int {{\varepsilon}_4}\sin
{{\theta }'} \,\text{d} {{\theta }'} \,\text{d}
{{\varphi }'}
\end{equation}
where the integration range is given by
${{\varepsilon}_4}>\varepsilon_\text{t}$. After integration over
$\theta'$ and $\varphi'$ we get
\begin{equation}
G=(1-F)\biggl[\varepsilon_\text{t}+\frac{{{(2 P+q)}^2}-4{Q^2}}{16m}\biggr]\,.
\end{equation}
The energy loss rate after integration over $P$ becomes
\begin{multline}
\dot{E}_\text{ev}=
\varepsilon_\text{t} \dot{N}_{\text{ev}}-
\frac{n_0^2 \text{e}^{-2\eta}}{128 \alpha^2 \sqrt{\beta m}} (\kB T)^4
\int_{0}^{\eta} \int_{0}^{\sqrt{8\kappa}}
\sigma(y)\, \sqrt{\eta-\kappa}\,
\Bigl( \eta - \kappa + \tfrac{U_0}{\kB T} \Bigr)\,y \\
\biggl\{ 256 y + 2 \sqrt{8\kappa-y^2}\,
\bigl[60+y^4-16y^2(\kappa-2)+64\kappa(\kappa-1)\bigr] \\
- 2\bigl\{ y^5 + 2y^4\sqrt{2\kappa} - 4y^3(4\kappa-3) 
- 16y^2\sqrt{2\kappa}\,(2\kappa-1) \\
+ y[68+32\kappa(2\kappa-3)]
+ 8 \sqrt{2\kappa}\,[15+16\kappa(\kappa-1)] \bigr\}
\exp {\bigl(\sqrt{2\kappa}\,y-\tfrac{1}{2}y^2\bigr)} \\
-\sqrt{\pi}\, \bigl( y^6 - 2y^4(12\kappa-7) + 4y^2[21+8\kappa(6\kappa-5)]
+ 8\{15-4\kappa[9+4\kappa(4\kappa-3)]\} \bigr) \\
\exp {\bigl( 2\kappa-\tfrac{1}{4}y^2 \bigr)}
\Bigl[
\erf {\Bigl( \sqrt{2\kappa-\tfrac{1}{4}y^2}\, \Bigr)} - 
\erf {\Bigl( \sqrt{2\kappa}-\tfrac{1}{2}y \Bigr)}
\Bigr]
\biggr\}\,\text{d} y \,\text{d} \kappa \,.
\label{bigEev}
\end{multline}

\begin{figure}
\includegraphics{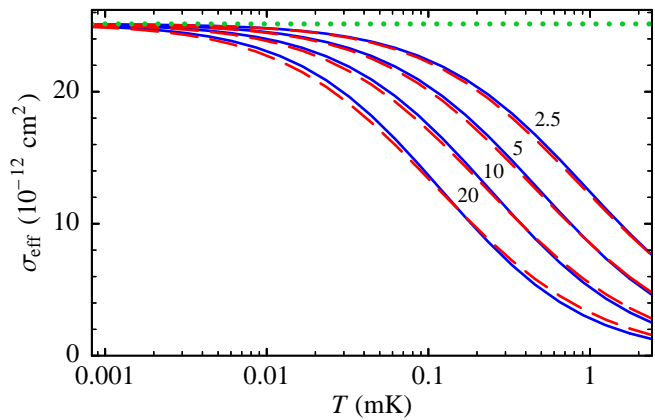}
\caption{Effective cross section as a function of temperature for
$m=4~\mathrm{u}$, $a=10~\mathrm{nm}$, $B_0=0.46~\mathrm{G}$ and four
values of $\eta$ (solid curves), together with the approximation
(dashed curves). The dotted line is the low-temperature limit.}
\label{theorysigma}
\end{figure}

When $\sigma$ is a constant, Eqs.~(\ref{bigNev}) and (\ref{bigEev})
reduce after integration to Eqs.~(\ref{Nev}) and (\ref{Eev}),
respectively. When $\sigma$ is energy-dependent, the expressions for
$\dot{N}_{\text{ev}}$ and $\dot{E}_{\text{ev}}$ have to be integrated
numerically. An effective cross section ${{\sigma }_{\text{eff}}}$ can
be introduced, which will give the correct value of
${\dot{N}_{\text{ev}}}$ if $\sigma ={{\sigma }_{\text{eff}}}$ in
Eq.~(\ref{Nev}). In the case that the cross section is given by
$\sigma = 8 \pi a^2 / (k^2 a^2 + 1)$, the effective cross section can
be approximated by a variation of Eq.~(\ref{colrate}):
\begin{equation}
\label{effsigma}
{{\sigma
}_{\text{eff}}}\approx 8\pi {a^{2 }} \bigl[\xi_{\text{ev}} - {{\xi
}_{\text{ev}}^2} {{\text{e} }^{{{\xi }_{\text{ev}}}}}\Gamma
(0,{{\xi }_{\text{ev}}})\bigr]\,,
\end{equation}
where
\begin{equation}
{{\xi }_{\text{ev}}}=\frac{3
}{\eta }\frac{{{\hbar }^2}}{m \kB T a^2}=\frac{3 {{\hbar
}^2}}{\varepsilon_\text{t}m a^2}\,.
\end{equation}
In this approximation, found by trial and error, the effective cross
section is only dependent on temperature via the trap
depth. Figure~\ref{theorysigma} shows ${{\sigma }_{\text{eff}}}$ and
its approximation as a function of temperature in our situation for
$a=10$~nm and several values of $\eta $. Apparently ${{\sigma
}_{\text{eff}}}(T)$ does not approach an asymptote at high $\eta $. In
general ($m<100$~u, $\abs{a}<100$~nm, $U_0/\muB<100$~G, $T<5$~mK,
$2<\eta <25$) the absolute difference between ${{\sigma
}_{\text{eff}}}$ and its approximation is less than 3\% of the
low-temperature limit $8\pi {a^2}$, shown as a dotted line in
Fig.~\ref{theorysigma}. The energy loss rate can still be calculated
with Eq.~(\ref{Eev}), even though the effective volumes are not
correct: the relative error in
${\dot{E}_{\text{ev}}}/{\dot{N}_{\text{ev}}}$ is less than 2\%.

\section{Experiment}
\label{secSweep}

\begin{figure}[!t]
\includegraphics{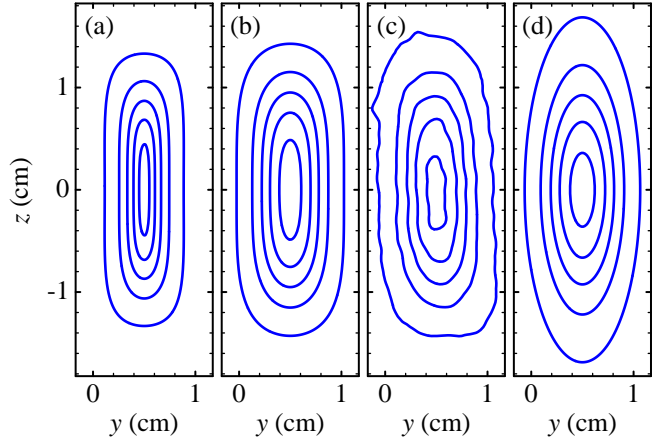}
\caption{The density distribution integrated in the
$x$-direction. From left to right: the theoretical distribution (a)
in the cloverleaf trap and (b) after 1~ms expansion, (c) the density
from the experimental absorption image (the negative logarithm of the
transmittance), and (d) a Gaussian fit to the experimental density
distribution. Contours at $10\%, 30\%, 50\%, 70\%$ and $90\%$ of the
maximum integrated density.}
\label{leafcontours}
\end{figure}

We have performed evaporative cooling experiments in a setup which is
described in detail by Herschbach {\em et
al.}~\cite{herschbach2003}. In short, an atomic beam from a DC
discharge source of metastable helium is collimated, deflected and
slowed in a traditional Zeeman slower. The atoms are trapped in a
magneto-optical trap (MOT), cooled in optical molasses, spin-polarized
by optical pumping, again trapped and finally compressed in a
cloverleaf magnetic trap, an example of an IQ trap.  The magnetic
field configuration is determined by $\alpha/(2\muB)=69$~G/cm,
$\beta/(2\muB)=13.9$~G/cm$^2$, $B_0=0.5$~G, and the trap depth is 47~G
(corresponding to 6.4~mK). The harmonic part is characterized by trap
frequencies $\omega_\rho/(2\pi) = 853$~Hz and $\omega_z/(2\pi) =
44$~Hz.  After compression a cloud of $\sim$10$^9$ atoms at a
temperature of $\sim$1~mK and a central density of
$\sim$10$^{10}$~cm$^{-3}$ is obtained. Temperature $T$ and a relative
measure of the number of atoms are determined by time-of-flight (TOF)
measurements with a microchannel plate (MCP) detector positioned 18~cm
from the trap center, whereas calibration of the number of atoms $N$
is performed by absorption imaging in the MOT. As the radial
confinement is much stronger than the axial confinement, the cloud is
elongated: the theoretical density while the atoms are still in the
trap has a FWHM of 0.2~cm horizontally and 1.4~cm vertically.  A
contour plot of the density, integrated in one horizontal dimension,
is shown in Fig.~\ref{leafcontours}(a). Absorption images are taken
1~ms after the trap has been switched off. To simulate the ballistic
expansion during this time, the theoretical density distribution is
convolved numerically with a Maxwell-Boltzmann velocity distribution
before the integration in one dimension, using a temperature of 1.2~mK
determined from a TOF measurement [Fig.~\ref{leafcontours}(b)]. No
fitting is required and the shape agrees well with the experimental
column density [Fig.~\ref{leafcontours}(c)]. In contrast, a Gaussian
fit to the experimental column density deviates significantly
[Fig.~\ref{leafcontours}(d)].
At lower temperatures the cloud becomes smaller and a larger part is
confined to the harmonic region of the trap. Therefore the density
distribution approaches a Gaussian shape during evaporative
cooling. However, only below 17~$\mu$K will the r.m.s.\ radii
calculated with a harmonic approximation of the trap deviate less
than 10\% from the exact solution.

\begin{figure}
\includegraphics{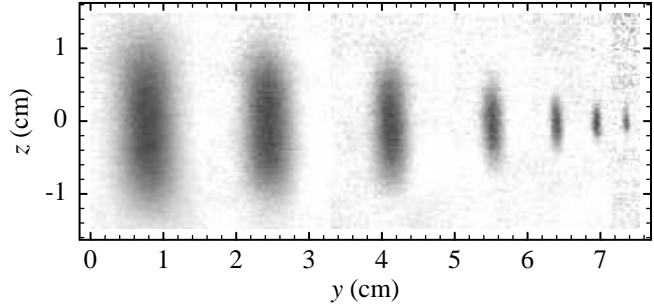} 
\caption{Collection of absorption images taken 1~ms after the
cloverleaf trap is switched off, at various stages during an rf
sweep. From left to right: just before rf is turned on, at 100~MHz,
65~MHz, 35~MHz, 15~MHz, 10~MHz, and 5~MHz.}
\label{rfimages}
\end{figure}

For evaporative cooling, the rf frequency starts at 120~MHz and
decreases exponentially with a time constant of 5~s, close to the
optimum found by maximizing the elastic collision rate per atom after
a sweep to 5~MHz. The progress of the temperature and number of atoms
is followed both in absorption imaging and with the MCP detector. A
series of images taken after 1~ms expansion is shown in
Fig.~\ref{rfimages}.
To emphasize the change in cloud size, maximum absorption within each
image is rendered black. The central absorption varies from 62\% in
the leftmost image to 14\% in the rightmost image. The smallest cloud
contains $9(4)\times {{10}^6}$ atoms at a temperature of
$23(3)$~$\mu$K.  Above 0.5~mK the radii are larger than expected,
especially in the horizontal plane, with a density that is half of the
predicted value. This may be explained by a time lag between a
change of trap depth and getting closer to steady state at a smaller
cloud size: the sweep has to be executed rather fast, due to a limited
trap lifetime of 12.5~s. For theoretical calculations the actual size
during the sweep is required. A first order correction of the modeled
density distribution is to multiply trap parameters $\alpha $ and
$\beta$ in Eq.~(\ref{leafgravpot}) with factors that are kept constant
during the sweep. These are ${f_{\alpha }}=0.74(9)$ and ${f_{\beta
}}=0.89(5)$, respectively, where the uncertainty includes the
variation in the measured radii and the temperature uncertainty.

\begin{figure}
\includegraphics{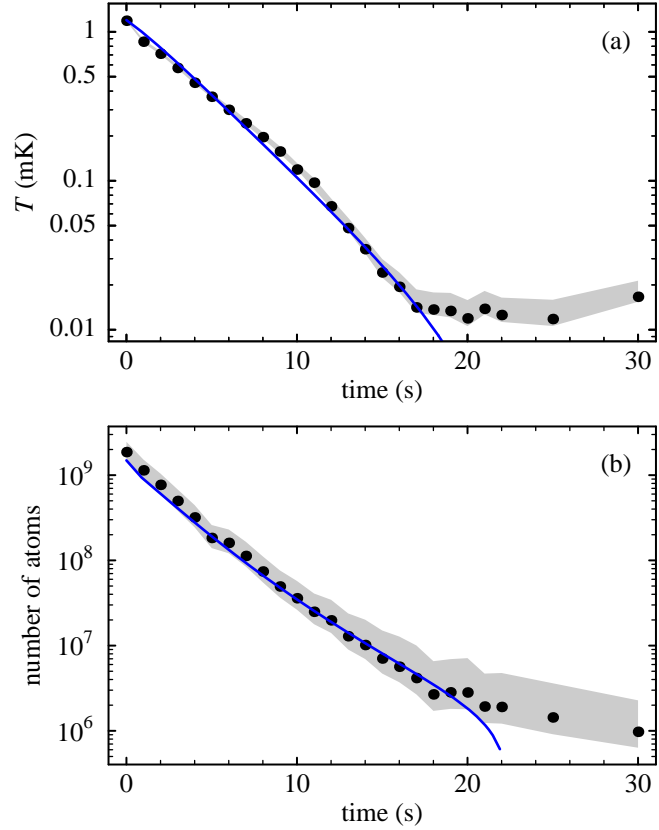}
\caption{(a) Temperature and (b) number of atoms as a function of rf time
 during a sweep. The gray bands show the possible values due to
 uncertainty in the detector calibration and in the temperature. The
 curve is a simulation with $a=10~\mathrm{nm}$.}
\label{TNsweep}
\end{figure}

Figure~\ref{TNsweep} shows the temperature and number of atoms as a
function of time since the start of the rf sweep. Each point is the
result of a TOF measurement after the sweep is interrupted by turning
off the rf power and the trap at the same time.
Error bars are given as a gray band, including a 20\% standard
deviation in the calibration of the number of atoms according to the
MCP and including the uncertainty in the temperature due to magnetic
field gradients between the trap and the MCP \cite{tolthesis}. The
indirect effect of the temperature error on the number of atoms is
also taken into account. The line is a theoretical calculation,
assuming the scattering length is 10~nm. The trap lifetime is measured
to be 12.5~s, the measured heating of 0.7~$\mu$K/s is neglected, and
the trap minimum is 0.5~G. Correction factors ${f_{\alpha }}$ and
${f_{\beta }}$ are included, so $\alpha / (2\muB) = 51$~G/cm and
$\beta / (2\muB) = 12.4$~G/cm$^2$. Due to an unexpected feature of the
rf generator, below 7~MHz the rf power drops, dwindling to zero below
4~MHz. This means the rf is effectively turned off after 17~s, the
temperature stays constant and the number of atoms goes down only due
to background collisions.

\begin{figure}
\includegraphics{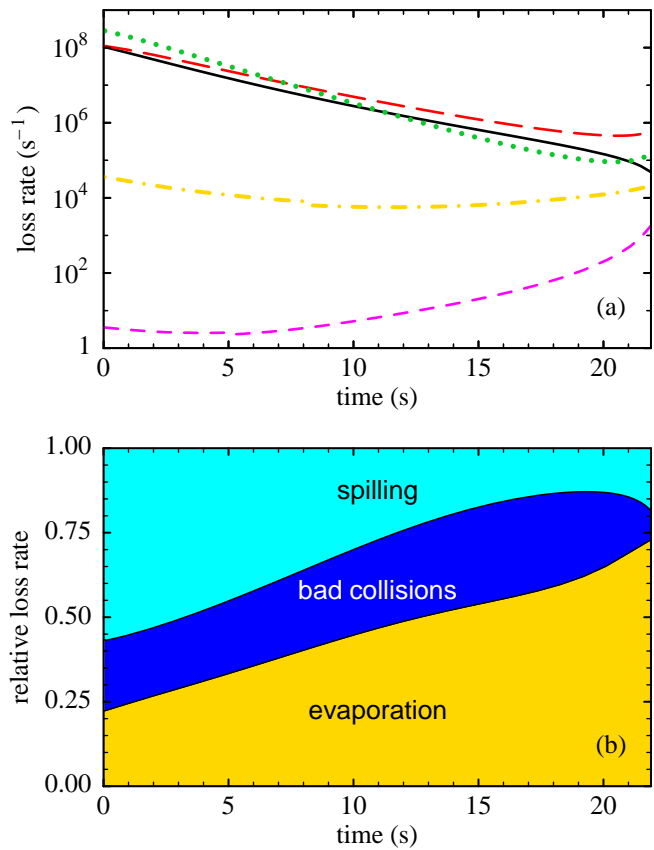}
\caption{(a) Absolute and (b) relative loss rates due to background
collisions $\dot{N}_\text{bgr}$ (solid curve), evaporation
$\dot{N}_\text{ev}$ (long-dashed curve), spilling $\dot{N}_\theta$
(dotted curve), two-body collisions $\dot{N}_\text{2b}$ (dash-dotted
curve) and three-body collisions $\dot{N}_\text{3b}$ (small-dashed
curve), as a function of sweep time. In (b) the three loss rates due
to bad inelastic collisions are summed. This is according to a
simulation with $a=10~\mathrm{nm}$.}
\label{sweeploss}
\end{figure}

The five loss rates used in the model are plotted as a function of
time in Fig.~\ref{sweeploss}, for the theoretical sweep with
$a=10$~nm. The contribution of two- and three-body collisions can be
neglected up to the moment BEC is reached, theoretically after
22~s. At the start of the sweep the dominant loss process is spilling,
because the frequency is decreased relatively fast. If the sweep would
take longer, spilling losses would be reduced in favor of evaporation
losses, but this would only be advantageous if the trap lifetime were
larger. Figure~\ref{sweeploss}(b) shows that an exponentially
decreasing rf frequency goes unnecessarily fast towards the end: the
spilling contribution increases in the last seconds.

\begin{figure}
\includegraphics{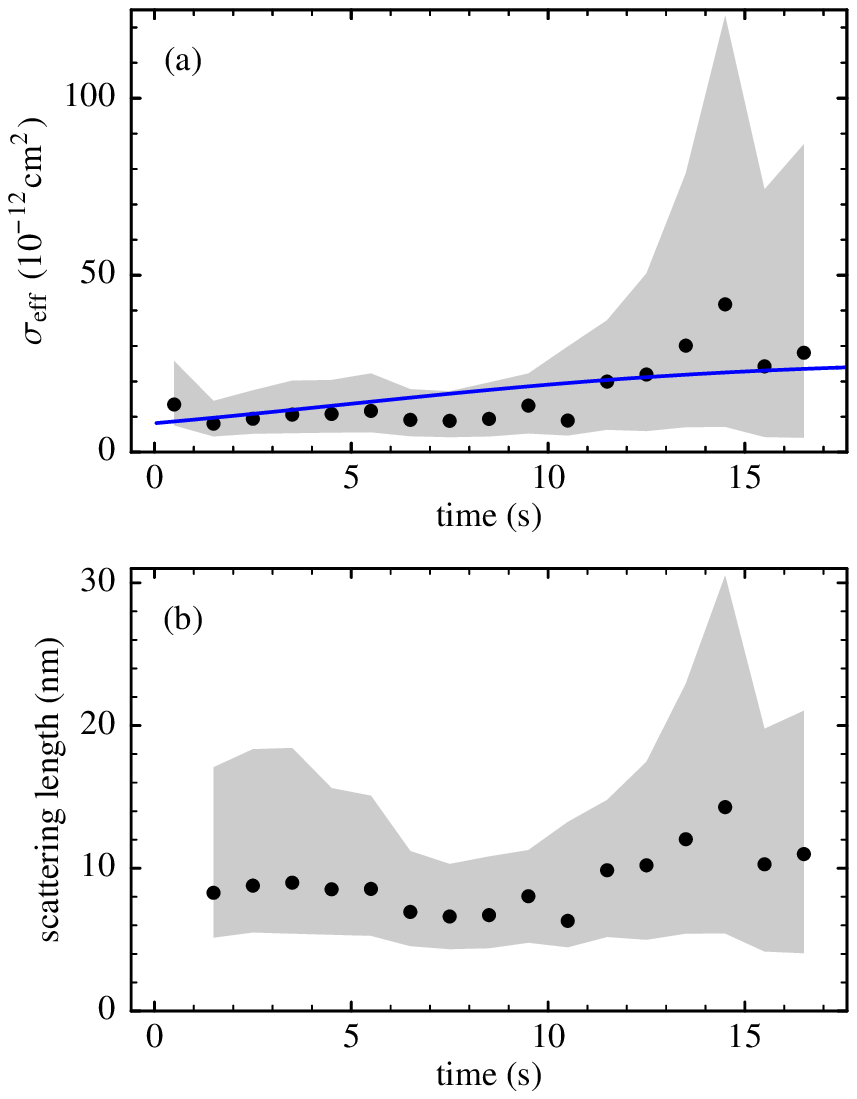}
\caption{(a) Effective cross section and (b) scattering length as a
function of sweep time. The curve is a simulation with
$a=10~\mathrm{nm}$, the gray bands indicate the error due to
uncertainty in the number of atoms, temperature and cloud size.}
\label{scatsweep}                         
\end{figure}

To find the scattering length best describing the data, one can
simulate sweeps for several values of $a$ starting from the
temperature and number of atoms in the beginning. However, deviations
early on influence the behavior of the curve at later times. For
instance, turning on the rf is not well modeled, as the experimental
temperature after 1~s is lower than expected [see
Fig.~\ref{TNsweep}(a)]. In addition, the temperature as a function of
time exhibits a slope change after 11~s, which is not seen in
simulations with any value of the scattering length. Therefore, small
sweeps are simulated for each interval between the data
points. Starting from the experimental situation at one point with a
constant $\sigma$ in Eq.~(\ref{Nev}), the temperature at the end of
the sweep is calculated. The values of $\sigma $ for which the end
temperature according to the model corresponds to the experimental
temperature are given in Fig.~\ref{scatsweep}(a). The curve is the
effective cross section according to Eq.~(\ref{effsigma}) for
$a=10$~nm. The gray band indicates the error due to the uncertainty in
the MCP calibration and the temperature (as in Fig.~\ref{TNsweep}), as
well as the uncertainty in the correction factor for the theoretical
cloud size.
The scattering length determined with the experimental effective cross
section and Eq.~(\ref{effsigma}) is given in
Fig.~\ref{scatsweep}(b). The point in the first interval is dropped,
because the effective cross section was unusually high. This in turn
is caused by the unexpectedly large drop in temperature after turning
on the rf power, which makes evaporation look very effective.  In the
last part of the sweep the uncertainty in the scattering length
becomes large due to the increasing uncertainty in the temperature
determination. However, the possible error in $a$ in the first half of
the sweep may be larger than given, because the evaporation process
seems not to behave entirely according to the model. Also, at the
start we have to rely more on the model anyway, as the scattering
length is determined only from evaporation, which then constitutes
just a quarter of the loss rate [Fig.~\ref{sweeploss}(b)]. Taking this
into account a scattering length $a=10(5)$~nm can be considered
as consistent with the experimental data.

\section{Summary and conclusions}
\label{secSummary}

To summarize, we have developed a model that allows simulation of
evaporative cooling experiments in situations where the elastic
scattering cross section depends on collision energy. The model is
used to simulate the number of atoms and temperature in an rf sweep
for metastable helium. From a comparison with the experiment an s-wave
scattering length $\abs{a}=10(5)$~nm is extracted. This value is a
factor of two smaller but within the experimental accuracy of previous
experiments; the BEC experiment of the Orsay group yields
$a=+20(10)$~nm \cite{robert, sirjean}, the result of the ENS group is
$a=+16(8)$~nm \cite{pereira2001, pereira2002}. Later experiments of
the ENS group are more consistent with a lower rather than a higher
value \cite{leduc}.  The scattering length can also be determined
theoretically, because the $^5\Sigma_\text{g}^+$ potential in which
spin-polarized atoms collide has been calculated. From the St\"arck
and Meyer potential \cite{starck} a scattering length $a=+8.3$~nm is
deduced. More recently, Gad\'ea, Leininger and Dickinson \cite{gadea}
calculated the short-range part of this potential more accurately and
combining their potential with the accurately determined long-range
potential of Yan and Babb \cite{yan} they determine
$a=+15.4$~nm. However, the value of the scattering length is very
sensitive to the way the short-range and long-range potential are
connected. An improved method of combining the two potentials results
in $a=+12.2^{+0}_{-4.1}$~nm \cite{dickinson}. Our experimental result
is in good agreement with this value.

\section*{Acknowledgments}

We gratefully acknowledge Norbert Herschbach for fruitful discussions
and the Foundation for Fundamental Research on Matter (FOM) for financial
support.


\bibliographystyle{apsrev}
\bibliography{shortnames,references}

\begin{thebibliography}{19}
\bibitem[{gam()}]{gammafunction}
\bibinfo{note}{Incomplete gamma function $\text{P}$ can be given in terms of
  incomplete gamma function $\Gamma $ as $\text{P}(a,z)=1-\Gamma (a,z)/\Gamma
  (a,0)= \int _{0}^{z}t^{a-1}\text{e}^{-t}\,\text{d}t \big/ \int _{0}^{\infty
  }t^{a-1}\text{e}^{-t}\,\text{d}t$}.
\end{thebibliography}
\end{document}